# Quasi-equilibrium relaxation of two identical quantum oscillators with arbitrary coupling strength

Illarion Dorofeyev[*]

Institute for Physics of Microstructures, Russian Academy of Sciences,

603950, GSP-105 Nizhny Novgorod, Russia

## Abstract

The paper deals with the problem of open systems out of equilibrium. An analytical expression for time-dependent density matrix of two arbitrary coupled identical quantum oscillators interacting with separate reservoirs is derived using path integral methods. A temporal behavior of spatial variances and of covariance from given initial values up to stationary values is investigated. It was shown that at comparatively low coupling strengths the asymptotic variances in the long-time limit achieve steady states despite on initial conditions. Stationary values of variances differ from the case of total equilibrium due to their coupling simultaneously with thermal reservoirs of different temperatures. The larger the difference in temperatures of thermal baths, the larger is the difference of the stationary values of variances of coupled oscillators comparing with values given by the fluctuation dissipation theorem. At strong couplings the variances have divergent character. Otherwise, in the weak coupling limit the asymptotic stationary variances are always in accordance with the fluctuation dissipation theorem despite of the difference in temperatures within the whole system.

*) E-mail: Illarion1955@mail.ru

# I. Introduction

The physics of open quantum systems covers a different set of phenomena ranging from nuclear to cosmic scales. It always deserves considerable attention because of a well known twofold inevitable effect of the environment on objects of interest. In this connection a quantum oscillator or an array of oscillators coupled to reservoirs is overall accepted models of open quantum systems. The models are often used to descibe the Brownian dynamics of selected particles coupled to a large number of harmonic oscillators, see for example [1-14]. Irreversibility in the dynamics of a quantum system interacting with a huge reservoir appears after reduction with respect to reservoir's variables. For the case of a harmonic oscillator with arbitrary damping and at arbitrary temperature an explicit expression for the time evolution of the density matrix when the system starts in a particular kind of pure state was derived and investigated in a seminal work [15] based on the path integral technique. It was shown that the spatial dispersion in the infinite time limit agrees with the fluctuation-disspation theorem (FDT). In order to study the transition to equilibrium state or to some intermediate stationary state of the system a problem for coupled oscillators interacting with heat baths in different approaches was considered in [16-21]. It was shown that an arbitrary initial state of a harmonic oscillator relaxes towards a uniquely determined stationary state. The evolution of quantum states of networks of quantum oscillators coupled with arbitrary external environments was analyzed in [22]. The emergence of thermodynamical laws in the long time regime and some constraints on the low frequency behavior of the environmental spectral densities were demonstrated.

Further, we quote results from papers describing bipartite continuous variable systems composed of two interacting oscillators closely relating to our study.

A study of the dynamics of entanglement and quantum discord between two oscillators coupled to a common environment was provided in papers [23-25]. Different phases of evolution including sudden death and revival of entanglement for different models of environments and for different models for the interaction between the system and reservoirs were described providing a characterization of the evolution for Ohmic, sub-Ohmic, and super-Ohmic non-Markovian environments.

In [26] the entanglement evolution of two harmonic oscillators under the influence of non-Markovian thermal environments was studied using non-Markovian master equations. They considered the cases of two separate and common baths and found that the dynamics of the quantum entanglement is sensitive to the initial states, the coupling between oscillators and the coupling to a common bath or to independent baths. In particular, it is been found that the entanglement can be sustained much longer when the two subsystems are coupled to a common bath than to independent baths.

Quantum decoherence of two coupled harmonic oscillators in a general environment at arbitrary temperature was studied in [27]. It was shown that the problem can be mapped into that of a single harmonic oscillator in a general environment plus a free harmonic oscillator. Some simplest cases of the entanglement dynamics were considered analytically and an analytical criterion for the finite-time disentanglement was derived under Markovian approximation.

An evolution of entanglement for a pair of coupled nonidentical oscillators in contact with an environment was studied in [28]. For cases of a common bath and of two separate baths, a full master equation is provided. The entanglement dynamics was analyzed as a function of the diversity between the oscillator's frequencies, a mutual coupling and also a correlation between occupation numbers. In case of separate baths at not very low temperatures the initial two-mode squeezed state becomes separable accompanying with a series of features. If the two oscillators share a common bath, the observation of asymptotic entanglement at relevant temperatures becomes possible. Their results also indicate that non-Markovian corrections at the weak-coupling approach can be observed only for a reduced subset of initial conditions.

The time evolution of quantum correlations of entangled two-mode states was examined in single-reservoir as well as two-resevoir models in [29]. They demonstrated that the properties of the stationary state may differ. Namely, in the two-reservoir model the initial entanglement is completely lost, and both modes are finally uncorrelated. In a common reservoir both modes interact indirectly via the coupling to the same bath variables. A separability criterion was derived.

A system of two coupled oscillators, each of them coupled to an independent reservoir is investigated in [30]. Their study has shown that a system of two particles each coupled to a heat bath at different temperatures, does not have a stationary state for all linear interactions. They have shown that if the baths are at different temperatures, then the interaction between the particles must be strong in order that there be a steady state entanglement. The same authors in [31] analyzed the same system in the high-temperature and weak coupling limits. The analytical solution of the non-rotating wave master equation is obtained. No thermal entanglement is found in the high-temperature regime. It was shown that in the weak coupling limit the system converges to an entangled non-equilibrium steady state.

Analytical expression for mean energy of interaction of two coupled oscillators within independent heat reservoirs of harmonic oscillators in a steady state regime was derived in [32]. Temporal dynamics of variances and covariance in the weak-coupling limit was studied in [33] based on path integral techniques. It was shown that despite on initial conditions the system of two weakly coupled oscillators in the infinite time limit agrees with the fluctuation dissipation theorem.

After reviewing the relative literature, we conclude that the study of evolution of interacting systems out of equilibrium at arbitrary coupling strengths from some initial state to some arbitrary states and realizability of the quasi-steady (quasi-equilibrium) states of the systems is mandatory and allows better understanding of complex phenomena within open systems.

The paper is devoted to analyse the temporal behavior of variances and covariance composing a density matrix of two arbitrary coupled identical oscillators within independent heat reservoirs. The main aim of this paper is to show the existence and reachability of quasi-equilibrium stationary states from given initial conditions of the system of two identical oscillators at some their coupling strengths interacting with separate baths of different temperatures.

The paper is organized as follows. In Sec.II we provide an expression for the reduced density matrix of two arbitrary coupled identical quantum oscillators interacting with different thermal baths. Limiting analytical formulas for the density matrix in various regimes with respect to the coupling constant and numerical study of a temporal behavior of variances and covariance from arbitrary initial states up to states in the infinite time limit are given in Sec.III. Our conclusions are given in Sec.IV.

## II. Problem statement and solution

As well as in our paper [33] we consider the system of two coupled oscillators where each of them is connected to a separate reservoir of independent oscillators. Both selected oscillators have equal masses $M$ and eigenfrequencies $\omega_0$ and $\lambda$ is their coupling constant ranging in value $-M\omega_0^2 \leq \lambda \leq M\omega_0^2$. In time $t < 0$ the whole system of oscillators is uncoupled. Then, the interactions are switched on in the time $t = 0$ and maintained during arbitrary time interval up to infinity. The problem is to find the time-dependent density matrix of two arbitrary coupled identical oscillators in any moment of time $t \geq 0$. Then, the reduced density matrix describing a propagation of two selected interacting oscillators from their coordinates $y_{1,2}$ to the coordinates $x_{1,2}$ during the time interval $t$ is presented as follows

$$\rho(x_1, x_2, y_1, y_2, t) = \int dx_1' dx_2' dy_1' dy_2'\, J(x_1, x_2, y_1, y_2, t; x_1', x_2', y_1', y_2', 0) \rho_A^{(1)}(x_1', y_1', 0) \rho_A^{(2)}(x_2', y_2', 0), \quad (1)$$

where $\rho_A^{(1)}(x_1', y_1', 0)$, $\rho_A^{(2)}(x_2', y_2', 0)$ are initially prepared density matrices of the two selected oscillators and $J(x_1, x_2, y_1, y_2, t; x_1', x_2', y_1', y_2', 0)$ is the propagator calculated for this problem in [33]. The final solution is obtained in new variables $X_{1,2} = x_{1,2} + y_{1,2}$, $\xi_{1,2} = x_{1,2} - y_{1,2}$. In these new variables the density matrix in Eq. (1) reads

$$\rho(X_{f1}, X_{f2}, \xi_{f1}, \xi_{f2}, t) = \int dX_{i1} dX_{i2} d\xi_{i1} d\xi_{i2}\, J(X_{f1}, X_{f2}, \xi_{f1}, \xi_{f2}, t; X_{i1}, X_{i2}, \xi_{i1}, \xi_{i2}, 0) \\ \times \rho_A^{(1)}(X_{i1}, \xi_{i1}, 0) \rho_A^{(2)}(X_{i2}, \xi_{i2}, 0) \quad (2)$$

where $X_{i1,i2} = X_{1,2}(0)$, $X_{f1,f2} = X_{1,2}(t)$ and $\xi_{i1,i2} = \xi_{1,2}(0)$, $\xi_{f1,f2} = \xi_{1,2}(t)$, and the propagating function in Eq.(2) can be represented in the following form

$$\begin{aligned}
J(X_{f1}, X_{f2}, \xi_{f1}, \xi_{f2}, t; X_{i1}, X_{i2}, \xi_{i1}, \xi_{i2}, 0) = \\
\tilde{C}_1 \tilde{C}_2 F_1^2(t) F_2^2(t) \exp\frac{i}{\hbar}\{\tilde{S}_{cl}^{(1)} + \tilde{S}_{cl}^{(2)} + \tilde{S}_{cl}^{(12)}\} \\
\times \exp-\frac{1}{\hbar}\{A_1(t)\xi_{f1}^2 + B_1(t)\xi_{f1}\xi_{i1} + C_1(t)\xi_{i1}^2\} \\
\times \exp-\frac{1}{\hbar}\{A_2(t)\xi_{f2}^2 + B_2(t)\xi_{f2}\xi_{i2} + C_2(t)\xi_{i2}^2\} \\
\times \exp-\frac{1}{\hbar}\{E_1(t)\xi_{i1}\xi_{i2} + E_2(t)\xi_{f2}\xi_{i1} + E_3(t)\xi_{f1}\xi_{i2} + E_4(t)\xi_{f1}\xi_{f2}\}
\end{aligned} \quad (3)$$

where expressions for the classical actions $\tilde{S}_{cl}^{(1)}$, $\tilde{S}_{cl}^{(2)}$ and $\tilde{S}_{cl}^{(12)}$ are formally the same as in Eqs.(38),(39) from [33].

Initial spatial variances of two oscillators in Eq.(2) are chosen in the Gaussian forms as follows

$$\begin{aligned}
\rho_A^{(1)}(X_{i1}, \xi_{i1}, 0) = (2\pi\sigma_{01}^2)^{-1/2} \exp\left[-(X_{i1}^2 + \xi_{i1}^2)/8\sigma_{01}^2\right], \\
\rho_A^{(2)}(X_{i2}, \xi_{i2}, 0) = (2\pi\sigma_{02}^2)^{-1/2} \exp\left[-(X_{i2}^2 + \xi_{i2}^2)/8\sigma_{02}^2\right],
\end{aligned} \quad (4)$$

where $\sigma_{01}^2$ and $\sigma_{02}^2$ are the initial dispersions of oscillators.

### III. Results and discussion.

#### A. The density matrix for coupled identical oscillators out of equilibrium.

After an integration of Eq.(2) with use of Eq.(4) the density matrix in case of identical coupled oscillators with arbitrary coupling strength is obtained also in Gaussian form

$$\rho(x_{f1}, x_{f2}, t) = \rho_{01}(t)\rho_{02}(t) \exp\left[-\frac{1}{2}\beta_{11}(t)x_{f1}^2 - \beta_{12}(t)x_{f1}x_{f2} - \frac{1}{2}\beta_{22}(t)x_{f2}^2\right], \qquad (5)$$

where

$$\beta_{11}(t, \lambda, T_1, T_2) = 2\left[\frac{(D_9 + D_9' + \Pi_6)^2}{\hbar(C_2 + \hbar a_2)} + \frac{Z_2^2}{\hbar^2 Z_1} - \frac{4e_6^2(C_2/\hbar + a_2)}{4\hbar a_2(C_2 + \hbar a_2) + (D_4' + \Pi_{16})^2} + \frac{Y_5^2}{\hbar^2 Y_1}\right], \qquad (6)$$

$$\beta_{22}(t, \lambda, T_1, T_2) = 2\left[\frac{(D_3' + \Pi_8)^2}{\hbar(C_2 + \hbar a_2)} + \frac{Z_3^2}{\hbar^2 Z_1} - \frac{4e_5^2(C_2/\hbar + a_2)}{4\hbar a_2(C_2 + \hbar a_2) + (D_4' + \Pi_{16})^2} + \frac{Y_4^2}{\hbar^2 Y_1}\right], \qquad (7)$$

$$\beta_{12}(t, \lambda, T_1, T_2) = \left[\frac{2(D_3' + \Pi_8)(D_9 + D_9' + \Pi_6)}{\hbar(C_2 + \hbar a_2)} + \frac{2Z_2 Z_3}{\hbar^2 Z_1} - \frac{8e_5 e_6(C_2/\hbar + a_2)}{4\hbar a_2(C_2 + \hbar a_2) + (D_4' + \Pi_{16})^2} + \frac{2Y_4 Y_5}{\hbar^2 Y_1}\right], \qquad (8)$$

$$\rho_{01}(t)\rho_{02}(t) = \frac{\tilde{C}_1 F_1^2(t)}{\sqrt{C_1(t) + \hbar a_1 + D_4^2(t)/4\hbar a_1}} \frac{\tilde{C}_2 F_2^2(t)}{\sqrt{C_2(t) + \hbar a_2 + D_4'^2(t)/4\hbar a_2}}, \qquad (9)$$

where $a_1 = 1/8\sigma_{01}^2$, $a_2 = 1/8\sigma_{02}^2$ and $F(t)$ is the wave function amplitude for the undamped case [15, 34] with renormalized eigenfrequencies of two oscillators due to their coupling [33]. The time dependent relevant functions $D, D', \Pi, C, E$ in Eqs.(6)-(9) for the problem of interaction of identical oscillators are presented in Appendices B and C.

It should be noted that in obtaining Eq.(5) from Eqs.(2)-(4) we put $X_f = 2x_f$ and $\xi_f = 0$ for simplicity as well as in [15].

From Eqs.(5)-(8) we obtain as usual [38] corresponding second moments

$$\sigma_1^2(t, \lambda, T_1, T_2) = \langle x_{f1}^2 \rangle = \frac{\beta_{22}(t, \lambda, T_1, T_2)}{\beta_{11}(t, \lambda, T_1, T_2)\beta_{22}(t, \lambda, T_1, T_2) - \beta_{12}^2(t, \lambda, T_1, T_2)}, \qquad (10)$$

$$\sigma_2^2(t, \lambda, T_1, T_2) = \langle x_{f2}^2 \rangle = \frac{\beta_{11}(t, \lambda, T_1, T_2)}{\beta_{11}(t, \lambda, T_1, T_2)\beta_{22}(t, \lambda, T_1, T_2) - \beta_{12}^2(t, \lambda, T_1, T_2)}, \qquad (11)$$

$$\beta_{12}^{-1}(t, \lambda, T_1, T_2) = \langle x_{f1} x_{f2} \rangle = \frac{\beta_{12}(t, \lambda, T_1, T_2)}{\beta_{12}^2(t, \lambda, T_1, T_2) - \beta_{11}(t, \lambda, T_1, T_2)\beta_{22}(t, \lambda, T_1, T_2)}, \qquad (12)$$

It should be emphasized that in general case of arbitrary coupling strength between oscillators the variances in Eqs.(6),(7) depend both on temperatures $T_1$ and $T_2$ differing to the case of a weak coupling [33]

### B. The limit of no coupling between oscillators.

In case of no coupling ($\lambda = 0$, $\Omega_1 = \Omega_2 = \omega_0$) all functions $\Pi = 0$ and it is easy to note from Appendix B that also $s_1 = s_3 = s_7 = s_{11}$, $s_2 = s_4 = s_{10} = s_{14}$, $s_5 = s_6 = s_8 = s_9 = s_{12} = s_{13}$. That is why $D_9 + D_9' = D_{10} + D_{10}' = D_{11} + D_{11}' = D_{12} + D_{12}' = 0$ and $e_3 = e_4 = e_6 = 0$, $e_5 = D_3' D_4' / 2(C_2 + \hbar a_2)$, $Z_1 = C_1/\hbar + a_1$, $Z_2 = D_3$, $Z_3 = 0$, $Z_6 = D_4$, $Y_1 = a_1 + Z_6^2 / 4\hbar^2 Z_1$, $Y_4 = 0$, $Y_5 = iZ_2 Z_6 / 2\hbar Z_1$. Moreover, because from (B22) it is follows that $f_1 = f_2 = f_3 = f_4$, $f_5 = f_6 = f_7 = f_8$, $f_9 = f_{11} = f_{13} = f_{15}$ and $f_{10} = f_{12} = f_{14} = f_{16}$ we have $E_1(t) = 0$, $C_1(t,T_1,T_2) = C_1(t,T_1)$, $C_2(t,T_1,T_2) = C_2(t,T_2)$ in (B20). As the result, we obtain $\sigma_{12}^2(t) = 0$ as it must be in this case, and Eqs.(10),(11) are transformed to the following expressions

$$\sigma_1^2(t,T_1) = \beta_{11}^{-1} = \frac{1}{2}\left\{\frac{D_3^2(t)}{\hbar[C_1(t,T_1) + \hbar a_1]}\left[1 - \frac{D_4^2(t)}{D_4^2(t) + 4\hbar a_1[C_1(t,T_1) + \hbar a_1]}\right]\right\}^{-1}, \quad (13)$$

$$\sigma_2^2(t,T_2) = \beta_{22}^{-1} = \frac{1}{2}\left\{\frac{D_3'^2(t)}{\hbar[C_2(t,T_2) + \hbar a_2]}\left[1 - \frac{D_4'^2(t)}{D_4'^2(t) + 4\hbar a_2[C_2(t,T_2) + \hbar a_2]}\right]\right\}^{-1}, \quad (14)$$

where

$$D_3(t) = D_3'(t) = (M/2)[-mn(b_1 + b_{13} + b_1' + b_{13}') + n(b_3 + b_{15} + b_3' + b_{15}')/2],$$
$$D_4(t) = D_4'(t) = (M/2)[m^2(b_1 + b_{13} + b_1' + b_{13}') - m(b_2 + b_3 + b_{15} + b_3' + b_{14}' + b_{15}')/2, \quad (15)$$
$$+ (b_4 + b_{16} + b_4' + b_{16}')/4]$$

$n(t) = \exp(\gamma t)/2\sin(\omega_0 t)$, $m(t) = \cot(\omega_0 t)/2$, the functions $b$ and $b'$ are from Eqs.(B11),(B12).

It is easy to see that Eqs.(13),(14) are equal to corresponding Eqs.(54),(55) from our paper [33]. Besides, these formulas are identically equal to Eq.(6.34) from [15] describing a relaxation of oscillator to an equilibrium state due to an interaction with a thermal bath.

### C. Temporal dependencies of variances and steady states of coupled oscillators.

The main goal of this work is to demonstrate relaxation peculiarities of coupled identical oscillators interacting with separate thermal baths kept with different temperatures. We recall that despite of his initial state one selected oscillator interacting with a thermostat reaches its equilibrium state and all his characteristics are given by the fluctuation dissipation theorem (FDT). For example, corresponding variance of the thermalized particle in steady state is always in accordance with the FDT [15]. We consider here different situations, namely when two oscillators are initially cold and characterized by their "natural" dispersions $\sigma_{01}^2 = \sigma_{02}^2 = \hbar/2M\omega_0$, and when the first oscillator is initially disturbed and has an arbitrary initial dispersion $\sigma_{01}^2 = 10(\hbar/2M\omega_0)$ but the second one has $\sigma_{02}^2 = \hbar/2M\omega_0$. For numerical calulations we have chosen the following parameters of oscillators: $M = 10^{-23} g$, $\omega_0 = 10^{13} rad/s$, and $\gamma = 0.01\omega_0$, relating to the typical characteristics for solid materials.

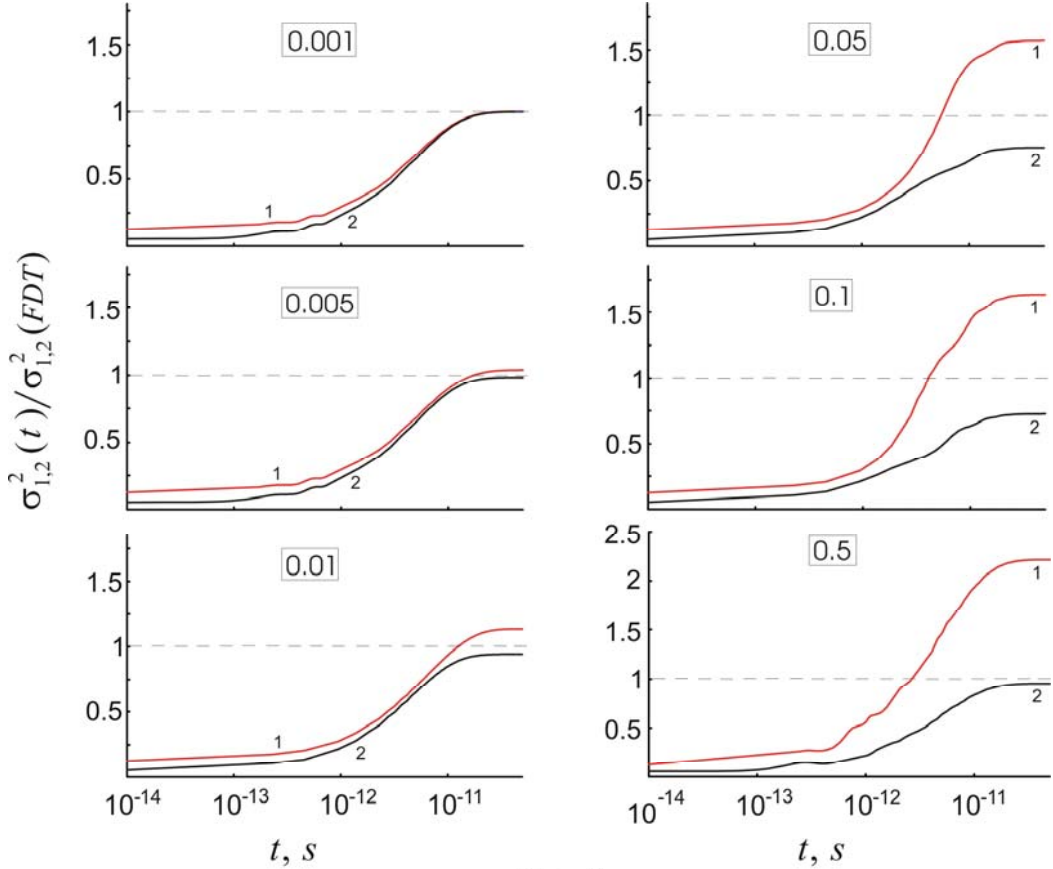

Fig. 1

Figure 1 illustrates the time-dependent dynamics of normalized variances $\sigma_{1,2}^2(t)/\sigma_{1,2}^2(FDT)$ in accordance with Eqs.(10),(11) in case of initially cold oscillators with dispersions $\sigma_{01}^2 = \sigma_{02}^2 = \hbar/2M\omega_0$. The insets in each picture denote the normalized coupling constant $\tilde{\lambda} = \lambda/M\omega_0^2$. Normalization was done to the variances of oscillators in equilibrium $\sigma_{1,2}^2(FDT)$ at $T_1 = 300K$ and $T_2 = 700K$ correspondingly using well known formula

$$\sigma_{1,2}^2(FDT) = \frac{\hbar}{\pi M} \int_0^\infty d\nu \, Coth\left(\frac{\hbar\nu}{2k_B T_{1,2}}\right) \frac{2\gamma\nu}{(\nu^2 - \omega_0^2)^2 + 4\gamma^2\nu^2}. \tag{16}$$

The pictures in Fig.1 illustrate the heating-up process of oscillators to some steady states at different coupling constants. The main point of this process is a deviation of stationary states of each oscillator from the states within a system in total equilibrium. The larger is the coupling constant, the larger is the deviation of the corresponding variance from its equilibrium value given by the FDT. The physical reason must be clear because from one side the first oscillator dispersion is determined by interaction with the first ("native") thermostat at $T_1 = 300K$ and additionally by smaller interaction with the second thermostat at $T_1 = 700K$ via the coupling with second oscillator, and vice versa. That is why the stationary variance of the first oscillator is a little bit larger, but the second stationary variance is a little bit smaller than their values in total equilibrium. It is possible to say judging from figure 1 that in range of some percentage the steady state variances are in accordance with FDT at $\tilde{\lambda} \leq 0.01$.

The second obvious point is almost monotonous dynamics both of identical oscillators initially prepared in natural states.

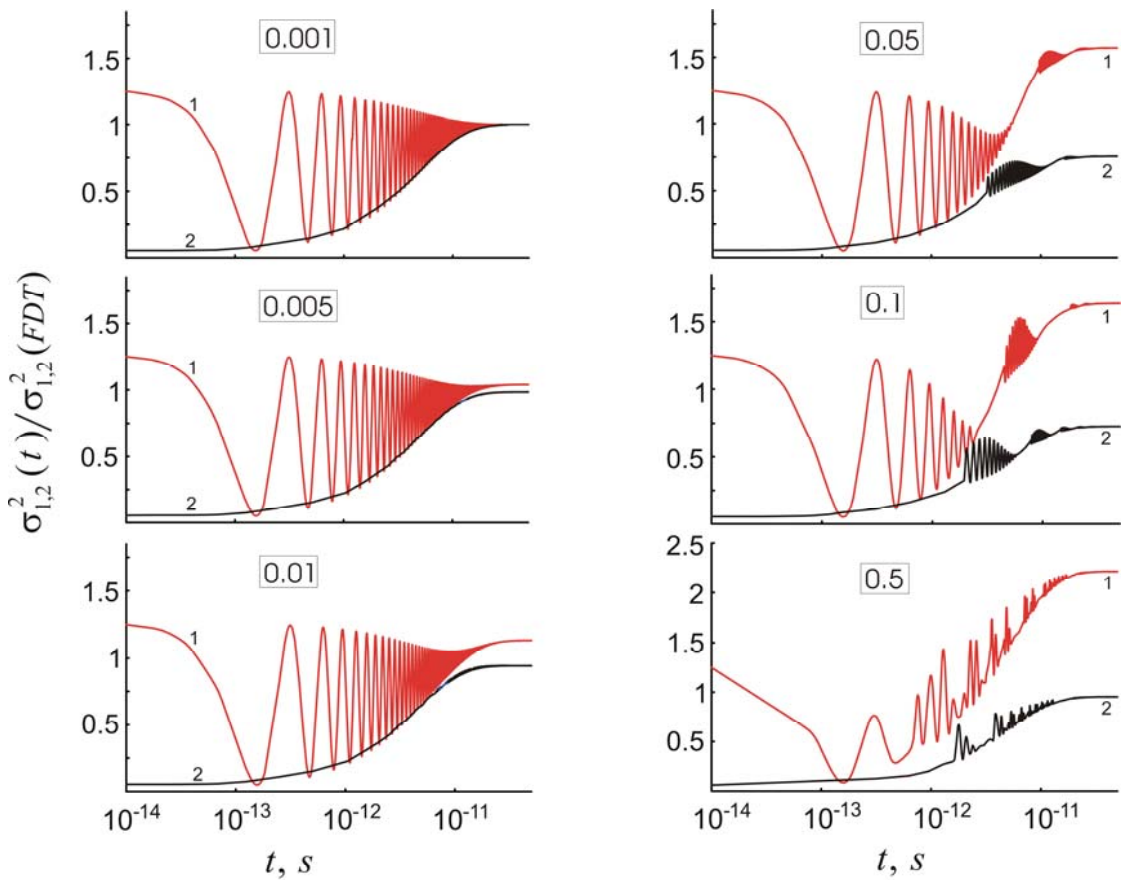

Fig.2

Fig.2 demonstrates the heating-up process of coupled oscillators to some steady states at the same set of coupling constants, but in case when the first oscillator is initially arbitrary disturbed with $\sigma_{01}^2 = 10(\hbar/2M\omega_0)$, but the second is cold with $\sigma_{02}^2 = \hbar/2M\omega_0$. The time-dependent dynamics of normalized variances $\sigma_{1,2}^2(t)/\sigma_{1,2}^2(FDT)$ calculated with use of Eqs.(10),(11) is shown in this figure. As well as in the previous figure, the insets in each picture denote the normalized coupling constant $\tilde{\lambda} = \lambda/M\omega_0^2$. It is seen that the initial excitation of the first oscillator yields in the non monotonic temporal behavior of the first dispersion at comparatively weak coupling constant (the left panel of the figure) and in the non monotonic behavior of both oscillators at comparatively strong coupling constant (the right panel). Moreover, at the right panel of the figure 2 we can to observe that the energy of excitation quasi-periodically spills over from one to other oscillator, demonstrating the well known effect of interaction of identical oscillators [35-37]. The oscillatory manner of the relaxation process of identical oscillators in case of our interest is a manifestation of complex interactions between these oscillators and their interactions with thermostats.

Based on results of our study shown in part in figures 1 and 2 we conclude that at comparatively strong coupling between oscillators their states in the long-time limit ($t \to \infty$) deviate from those following from the case of total equilibrium described by the fluctuation dissipation theorem.

To describe the deviation of normalized variances from the unit in steady states, which in accordance with rigid requirements of the FDT, we calculate the normalized dispersions $\tilde{\sigma}_{1,2}^2(t=\infty) = \sigma_{1,2}^2(t=\infty)/\sigma_{1,2}^2(FDT)$ and covariances $\tilde{\beta}_{12}^{-1}(t=\infty) = \beta_{12}^{-1}(t=\infty)/\sqrt{\sigma_1^2(FDT)\sigma_2^2(FDT)}$ in the long-time limit of coupled oscillators interacting with different thermostats kept with identical $T_1 = T_2 = 300K$ and different temperatures $T_1 = 300K$ and $T_2 = 700K$ correspondingly.

Figure 3 exemplifies $\tilde{\sigma}_{1,2}^2(t=\infty)$ and $\tilde{\beta}_{12}^{-1}(t=\infty)$ versus the normalized coupling constant $\tilde{\lambda}$ in case of total equilibrium -a), and in case of different temperatures –b). The principal common peculiarity in these both figures is the divergent behavior of $\tilde{\sigma}_{1,2}^2(t=\infty)$ and $\tilde{\beta}_{12}^{-1}(t=\infty)$ at $\tilde{\lambda} \to 1$. Also, in Fig.3b) we observe the splitting of curves 1 and 2 in case of different temperatures of thermostats. It is clearly seen that the distingushable splitting occures at $\tilde{\lambda} \geq 0.001$ followed by the increase at larger values of $\tilde{\lambda}$. In Fig.3c) we show covariances $\tilde{\beta}_{12}^{-1}(t=\infty)$ at small coupling strengths for cases of different (curve 1) and identical (curve2) temperatures. In Fig.3c) we presented also the results of calculations of covariances in accordance with papers [30, 31] where the formulas were derived at $k_B T \gg \hbar\omega_0$ in a steady state regime at weakly coupling approach $\tilde{\lambda} \ll 1$. In our example we have $k_B T \geq \hbar\omega_0$. Nevertheless, we observe very good qualitative and quantitative coincidence of the results.

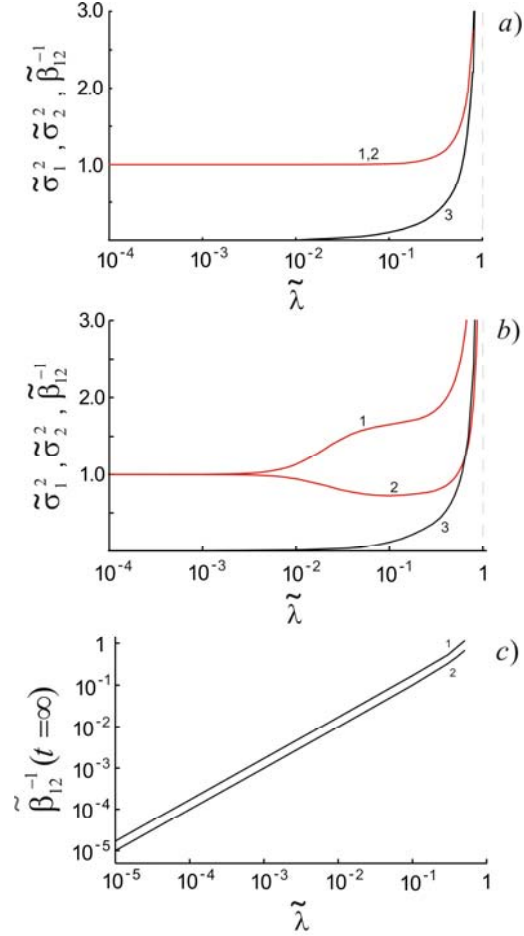

Fig.3

Here we want to conclude that in a system out of equilibrium the variances of coupled oscillators are in agreement with the FDT only at comparatevly small coupling constants. At large enough couplings and at large the difference in temperatures $T_1 - T_2$ the deviations of variances from their equilibrium values may be quite significant.

To clarify the observed divergent behavior of variances and covariances at $\lambda \to m\omega_0^2$ we consider an appropriate transformation of Hamiltonian for two bilinear coupling oscillators when each of them, in its turn, is bilinear coupled to separate reservoirs of oscillators. Corresponding Hamiltonian is written as follows

$$H = p_1^2/2M_1 + M_1\omega_{01}^2 x_1^2/2 + p_2^2/2M_2 + M_2\omega_{02}^2 x_2^2/2 - \lambda x_1 x_2 +$$
$$+\sum_{j=1}^{N_1}\left[p_j^2/2m_j + m_j\omega_j^2(q_j - x_1)^2/2\right] + \sum_{k=1}^{N_2}\left[p_k^2/2m_k + m_k\omega_k^2(q_k - x_2)^2/2\right], \quad (17)$$

where $x_{1,2}$, $p_{1,2}$, $M_{1,2}$, $\omega_{01,02}$ are the coordinates, momenta, masses and eigenfrequencies of the selected oscillators, $\lambda$ is the coupling constant, $q_j, p_j, \omega_j, m_j$ and $q_k, p_k, \omega_k, m_k$ are the coordinates, momenta, eigenfrequencies and masses of bath's oscillators.

After changing the variables $X = x_1 + x_2$ and $Y = x_1 - x_2$ we have instead of Eq.(17) the following expression

$$\begin{aligned}H &= \left[(M_1+M_2)/4\right]\dot{X}^2/2 + \left[(M_1\omega_{01}^2 + M_2\omega_{02}^2)/4 + \lambda/2\right]X^2/2 \\ &+ \left[(M_1+M_2)/4\right]\dot{Y}^2/2 + \left[(M_1\omega_{01}^2 + M_2\omega_{02}^2)/4 - \lambda/2\right]Y^2/2 \\ &+ \left[(M_1-M_2)/4\right]\dot{X}\dot{Y} + \left[(M_1\omega_{01}^2 - M_2\omega_{02}^2)/4\right]XY \\ &+ \sum_{j=1}^{N_1}\left\{p_j^2/2m_j + m_j\omega_j^2\left[q_j - (X+Y)/2\right]^2/2\right\} \\ &+ \sum_{k=1}^{N_2}\left\{p_k^2/2m_k + m_k\omega_k^2\left[q_k - (X-Y)/2\right]^2/2\right\}\end{aligned} \quad (18)$$

Then in case of two identical oscillators $M_1 = M_2 = M$, $\omega_{01} = \omega_{02} = \omega_0$, we obtain

$$\begin{aligned}H &= P_1^2/2M' + M'(\omega_0^2 + \lambda/M)X^2/2 \\ &+ P_2^2/2M' + M'(\omega_0^2 - \lambda/M)Y^2/2 \\ &+ \sum_{j=1}^{N_1}\left\{p_j^2/2m_j + m_j\omega_j^2\left[q_j - (X+Y)/2\right]^2/2\right\}, \\ &+ \sum_{k=1}^{N_2}\left\{p_k^2/2m_k + m_k\omega_k^2\left[q_k - (X-Y)/2\right]^2/2\right\}\end{aligned} \quad (19)$$

where $P_1 = M'\dot{X}$ and $P_2 = M'\dot{Y}$, $M' = M/2$.

Thus, we have in Eq.(19) the Hamiltonian for two uncoupled oscillators with the new masses $M' = M/2$ and new eigenfrequencies $\omega_0^2 \pm \lambda/M$ for the symmetric and antisymmetric modes interacting only via the thermostats. Moreover, when the coupling constant $\lambda \to M\omega_0^2$ the Hamiltonian tends to the Hamiltonian for a free particle and independent oscillator. That is why, due to the fictious free particle appearing at this condition, the dispersions are divergent at $\lambda = M\omega_0^2$ in Fig.3 because $\langle x_{free}^2\rangle = 2Dt$, $D = k_B T/M\gamma$ in this case. The same conclusion was done previously in [30] by analysing motion equations.

The situation is illustrated in Fig.4. We note here that in [27] it was shown that the two oscillator model can be effectively mapped into that of a single harmonic oscillator in a general environment plus a free harmonic oscillator.

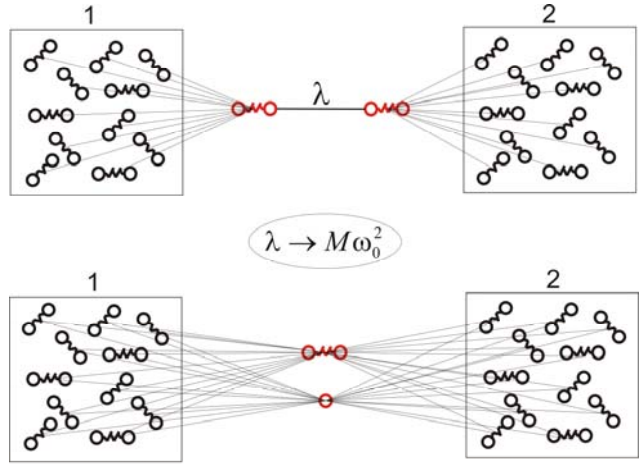

Fig.4

Finally, in order to illustrate the splitting of two variances due to the difference in temperatures in Fig.5 we present the temperature dependence of normalized variances $\tilde{\sigma}_{1,2}^2(t=\infty)$ in quasistationary states versus $T_2$ at fixed $T_1 = 300K$. Two pairs of curves were drawn at comparatively small values of the coupling constant. It should be emphasized that the variances in Eqs.(10),(11) are even functions but the covariance in Eq.(12) is odd function with respect to the coupling constant $\lambda$.

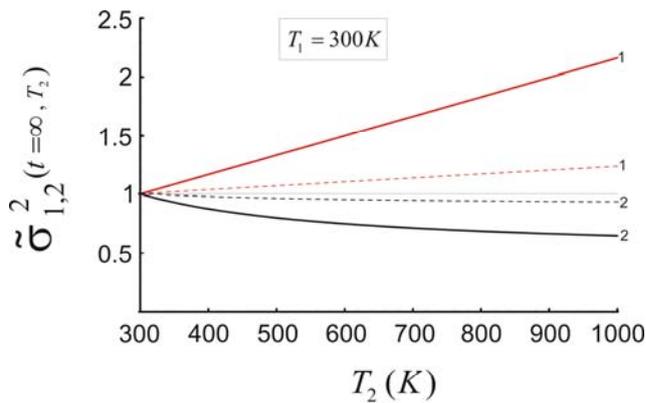

Fig.5

Thus, we demonstrated that the quasi-equilibrium states of two coupled identical oscillators interacting with separate baths kept at different temperatures are reachable at comparatively small coupling strengths ($\tilde{\lambda} \leq 0.01$), see for instance Fig.3b). Besides, at larger coupling strengths ($\tilde{\lambda} > 0.01$) the variances and covariances may reach stationary values different to the corresponding case of equilibrium. At the condition $\tilde{\lambda} \to 1$ these characteristics are divergent.

## IV. Conclusion.

Our paper aims to analyze the temporal dynamics and reachibility of quasi-steady states of pair coupled oscillators interacting with different thermostats. This is a simplest example for the problem of relaxation of open systems out of equilibrium. An analytical expression for time-dependent density matrix in this case is derived using path integral methods. The temporal dependencies of spatial variances and covariances from given initial values up to stationary values are investigated. It was shown that at comparatively low coupling strengths asymptotic variances in the long-time limit achieve steady states despite on initial conditions. Stationary level values of variances differ from the case of total equilibrium due to effective coupling of oscillators simultaneously with thermal reservoirs of different temperatures. The larger the difference in temperatures of thermal baths, the larger is the difference of the stationary values of variances of coupled oscillators comparing with values given by the fluctuation dissipation theorem. At strong couplings the dispersions of oscillators have divergent character. Corresponding clarification is proposed based on transformation of the initial Hamiltonian to the Hamiltonian containing the term for a free particle. In the weak coupling limit [33] the asymptotic stationary variances are always in accordance with the fluctuation dissipation theorem despite of the difference in temperatures (reasonable for solids, of course) within the whole system.

## Acknowledgements

The author thanks everybody who likes physics.

## Appendix A. Solution to the classical motion equations.

It is shown in our work [33] that the Lagrangian for the problem of two interacting oscillators is as follows

$$\begin{aligned}\mathcal{L} = &\, M_1 \dot{x}_1^2/2 - M_1 \dot{y}_1^2/2 + M_2 \dot{x}_2^2/2 - M_2 \dot{y}_2^2/2 \\ &- M_1 \omega_{01}^2 x_1^2/2 + M_1 \omega_{01}^2 y_1^2/2 - M_2 \omega_{02}^2 x_2^2/2 + M_2 \omega_{02}^2 y_2^2/2 \\ &- M_1 \gamma_1 [x_1 \dot{x}_1 - y_1 \dot{y}_1 + x_1 \dot{y}_1 - y_1 \dot{x}_1] - M_2 \gamma_2 [x_2 \dot{x}_2 - y_2 \dot{y}_2 + x_2 \dot{y}_2 - y_2 \dot{x}_2] \\ &+ \lambda x_1 x_2 - \lambda y_1 y_2\end{aligned} \quad (A1)$$

In the new variables $X_{1,2} = x_{1,2} + y_{1,2}$, $\xi_{1,2} = x_{1,2} - y_{1,2}$ the Lagrangian in Eq. (A1) reads

$$\begin{aligned}\mathcal{L} = &\, M_1 \dot{X}_1 \dot{\xi}_1/2 - M_1 \omega_{01}^2 X_1 \xi_1/2 - M_1 \gamma_1 \dot{X}_1 \xi_1 \\ &+ M_2 \dot{X}_2 \dot{\xi}_2/2 - M_2 \omega_{02}^2 X_2 \xi_2/2 - M_2 \gamma_2 \dot{X}_2 \xi_2 \\ &+ (\lambda/2)(X_1 \xi_2 + X_2 \xi_1)\end{aligned} \quad (A2)$$

As well as in [15, 34] we represent the paths as the sums $X_{1,2} = \tilde{X}_{1,2} + X'_{1,2}$, $\xi_{1,2} = \tilde{\xi}_{1,2} + \xi'_{1,2}$, explicitly selecting classical paths $\tilde{X}_{1,2}$, $\tilde{\xi}_{1,2}$ and fluctuating parts $X'_{1,2}$, $\xi'_{1,2}$ with boundary conditions $X'_{1,2}(0) = X'_{1,2}(t) = 0$, $\xi'_{1,2}(0) = \xi'_{1,2}(t) = 0$. The Lagrangian in Eq.(A2) becomes

$$\begin{aligned}\mathcal{L} \equiv \tilde{\mathcal{L}} + \mathcal{L}' = &\, M_1 \dot{\tilde{X}}_1 \dot{\tilde{\xi}}_1/2 - M_1 \omega_{01}^2 \tilde{X}_1 \tilde{\xi}_1/2 - M_1 \gamma_1 \dot{\tilde{X}}_1 \tilde{\xi}_1 \\ &+ M_2 \dot{\tilde{X}}_2 \dot{\tilde{\xi}}_2/2 - M_2 \omega_{02}^2 \tilde{X}_2 \tilde{\xi}_2/2 - M_2 \gamma_2 \dot{\tilde{X}}_2 \tilde{\xi}_2 + (\lambda/2)(\tilde{X}_1 \tilde{\xi}_2 + \tilde{X}_2 \tilde{\xi}_1) \\ &+ M_1 \dot{X}'_1 \dot{\xi}'_1/2 - M_1 \omega_{01}^2 X'_1 \xi'_1/2 - M_1 \gamma_1 \dot{X}'_1 \xi'_1 \\ &+ M_2 \dot{X}'_2 \dot{\xi}'_2/2 - M_2 \omega_{02}^2 X'_2 \xi'_2/2 - M_2 \gamma_2 \dot{X}'_2 \xi'_2 + (\lambda/2)(X'_1 \xi'_2 + X'_2 \xi'_1)\end{aligned} \quad (A3)$$

From the Lagrange equations of motion

$$\frac{d}{dt}\frac{\partial \mathcal{L}}{\partial \dot{\tilde{X}}_i} - \frac{\partial \mathcal{L}}{\partial \tilde{X}_i} = 0, \quad \frac{d}{dt}\frac{\partial \mathcal{L}}{\partial \dot{\tilde{\xi}}_i} - \frac{\partial \mathcal{L}}{\partial \tilde{\xi}_i} = 0, \quad (i = 1, 2), \quad (A4)$$

we have two pairs of coupled equations

$$\begin{cases} \ddot{\tilde{X}}_1 + 2\gamma_1 \dot{\tilde{X}}_1 + \omega_{01}^2 \tilde{X}_1 = (\lambda/M_1)\tilde{X}_2 \\ \ddot{\tilde{X}}_2 + 2\gamma_2 \dot{\tilde{X}}_2 + \omega_{02}^2 \tilde{X}_2 = (\lambda/M_2)\tilde{X}_1 \end{cases}, \quad (A5)$$

$$\begin{cases} \ddot{\tilde{\xi}}_1 - 2\gamma_1 \dot{\tilde{\xi}}_1 + \omega_{01}^2 \tilde{\xi}_1 = (\lambda/M_1)\tilde{\xi}_2 \\ \ddot{\tilde{\xi}}_2 - 2\gamma_2 \dot{\tilde{\xi}}_2 + \omega_{02}^2 \tilde{\xi}_2 = (\lambda/M_2)\tilde{\xi}_1 \end{cases}, \quad (A6)$$

We can seek a solution of the equations using various methods [35-37], for example in the following general form $X_{1,2} = A_{1,2} \exp(\varepsilon \tau)$. Substitution of this form into Eq.(A5) and taking $\varepsilon = i\omega - \delta$ in a determinant equation yield

$$\omega^4 - (\omega_{01}^2 + \omega_{02}^2 + \Delta_1)\omega^2 + \omega_{01}^2\omega_{02}^2 - \lambda^2/M_1M_2 + \Delta_2 = 0, \qquad (A7)$$

$$\begin{aligned}(\omega^2 - \omega_{02}^2)(\gamma_1 - \delta) + (\omega^2 - \omega_{01}^2)(\gamma_2 - \delta) \\ -(\gamma_1 - \delta)(2\gamma_2\delta - \delta^2) - (\gamma_2 - \delta)(2\gamma_1\delta - \delta^2) = 0\end{aligned}, \qquad (A8)$$

where $\Delta_1 = 4(\gamma_1 - \delta)(\gamma_2 - \delta) - (2\gamma_1\delta - \delta^2) - (2\gamma_2\delta - \delta^2)$,

$\Delta_2 = \omega_{01}^2 + \omega_{02}^2 + (2\gamma_1\delta - \delta^2)(2\gamma_2\delta - \delta^2) - (2\gamma_1\delta - \delta^2)\omega_{02}^2 - (2\gamma_2\delta - \delta^2)\omega_{01}^2$.

The roots of the equation (A7) are

$$\Omega_{1,2}^2 = (\omega_{01}^2 + \omega_{02}^2 + \Delta_1)/2 \mp \sqrt{(\omega_{01}^2 + \omega_{02}^2 + \Delta_1)^2/4 - \Delta_2 + \lambda^2/M_1M_2}. \qquad (A9)$$

From Eq.(A8) we obtain

$$\delta = \frac{(\omega^2 - \omega_{02}^2)\gamma_1 + (\omega^2 - \omega_{01}^2)\gamma_2}{(\omega^2 - \omega_{02}^2) + (\omega^2 - \omega_{01}^2)}, \qquad (A10)$$

where we neglected In Eq.(A8) by terms proportional to the third power with respect to dissipative parameters.

In case of pair identical oscillators ($\omega_{01,02} = \omega_0$, $\gamma_1 = \gamma_2 = \gamma$, $M_1 = M_2 = M$) we have from Eq.(A9)

$$\Omega_{1,2}^2 = \omega_0^2 - \gamma^2 \mp \lambda/M, \qquad (A11)$$

where we ordered the new eigenfrequencies as follows $\Omega_1 < \omega_0 < \Omega_2$.

After substitution of $\Omega_{1,2}^2$ into Eq.(A10) we obtain $\delta_{1,2}$ for the first and second modes.

Besides, from the Eq.(A5) we have two important ratios

$$\begin{aligned}\frac{A_2}{A_1} \equiv r_1(\Omega_1) = |r_1(\Omega_1)|\exp(i\kappa_1) = \frac{(\omega_{01}^2 - 2\gamma_1\delta + \delta^2) - \Omega_1^2 + 2i\Omega_1(\gamma_1 - \delta)}{\lambda/M_1}, \\ \frac{A_1}{A_2} \equiv r_2(\Omega_2) = |r_2(\Omega_2)|\exp(i\kappa_2) = \frac{(\omega_{02}^2 - 2\gamma_2\delta + \delta^2) - \Omega_2^2 + 2i\Omega_2(\gamma_2 - \delta)}{\lambda/M_2},\end{aligned} \qquad (A12)$$

which determine a relative contribution to the dynamics of oscillators from each eigenmode.

The loss angles in Eq.(A12) are determined as follows

$$\tan(\kappa_1) = \frac{2\Omega_1(\gamma_1 - \delta)}{\omega_{01}^2 - 2\gamma_1\delta + \delta^2 - \Omega_1^2}, \quad \tan(\kappa_2) = \frac{2\Omega_2(\gamma_2 - \delta)}{\omega_{02}^2 - 2\gamma_2\delta + \delta^2 - \Omega_2^2}. \qquad (A13)$$

It follows that the coefficients $r_{1,2}$ are complex in general, but in case $\gamma_{1,2} = \gamma$ we have from Eq.(A10) $\delta_{1,2} = \gamma$ and pure real $r_{1,2}$ in Eqs.(A12) due to Eq.(A13). This allows simplifying the analysis, while still keeping within a framework which provides consideration for an arbitrary coupling between oscillators. Thus, we consider the pure real coefficients

$$r_1 = \frac{\omega_{01}^2 - \gamma^2 - \Omega_1^2}{\lambda/M_1}, \quad r_2 = \frac{\omega_{02}^2 - \gamma^2 - \Omega_2^2}{\lambda/M_2}. \tag{A14}$$

Taking into account Eq.(A11) we have $r_1 = 1$ and $r_2 = -1$ in case of identical oscillators.

To construct a general solution we follow the prescription as recommended, for example in [35-37]

$$\begin{aligned}\tilde{X}_1(\tau) &= A_1 \sin(\Omega_1 \tau + \varphi_1)\exp(-\delta_1 \tau) + r_2 A_2 \sin(\Omega_2 \tau + \varphi_2)\exp(-\delta_2 \tau),\\ \tilde{X}_2(\tau) &= r_1 A_1 \sin(\Omega_1 \tau + \varphi_1)\exp(-\delta_1 \tau) + A_2 \sin(\Omega_2 \tau + \varphi_2)\exp(-\delta_2 \tau)\end{aligned} \tag{A15}$$

where the coefficients $r_{1,2}$ from Eq.(A14). The above equations make clear their physical sense, they determine a relative contribution from the first or second mode to the dynamics of each oscillators.

Solutions of the system (A1) in the general form at $r_1 = 1$ and $r_2 = -1$ satisfying the conditions $X_{1,2}(0) = X_{i1,2}$ and $X_{1,2}(t) = X_{f1,2}$ are as follows

$$\begin{aligned}\tilde{X}_1(\tau) &= \left[\frac{X_{f1} + X_{f2}}{2\sin(\Omega_1 t)}\exp(\gamma t) - \cot(\Omega_1 t)\frac{X_{i1} + X_{i2}}{2}\right]\sin(\Omega_1 \tau)\exp(-\gamma \tau) +\\ &+ \frac{X_{i1} + X_{i2}}{2}\cos(\Omega_1 \tau)\exp(-\gamma \tau) -\\ &- \left[\frac{X_{f2} - X_{f1}}{2\sin(\Omega_2 t)}\exp(\gamma t) - \cot(\Omega_2 t)\frac{X_{i2} - X_{i1}}{2}\right]\sin(\Omega_2 \tau)\exp(-\gamma \tau) -\\ &- \frac{X_{i2} - X_{i1}}{2}\cos(\Omega_2 \tau)\exp(-\gamma \tau)\end{aligned} \tag{A16}$$

$$\begin{aligned}\tilde{X}_2(\tau) &= \left[\frac{X_{f1} + X_{f2}}{2\sin(\Omega_1 t)}\exp(\gamma t) - \cot(\Omega_1 t)\frac{X_{i1} + X_{i2}}{2}\right]\sin(\Omega_1 \tau)\exp(-\gamma \tau) +\\ &+ \frac{X_{i1} + X_{i2}}{2}\cos(\Omega_1 \tau)\exp(-\gamma \tau) +\\ &+ \left[\frac{X_{f2} - X_{f1}}{2\sin(\Omega_2 t)}\exp(\gamma t) - \cot(\Omega_2 t)\frac{X_{i2} - X_{i1}}{2}\right]\sin(\Omega_2 \tau)\exp(-\gamma \tau) +\\ &+ \frac{X_{i2} - X_{i1}}{2}\cos(\Omega_2 \tau)\exp(-\gamma \tau)\end{aligned} \tag{A17}$$

By the same way we obtained solutions of the system (A6) satisfying the conditions $\xi_{1,2}(0) = \xi_{i1,2}$ and $\xi_{1,2}(t) = \xi_{f1,2}$

$$\tilde{\xi}_1(\tau) = \left[\frac{\xi_{f1}+\xi_{f2}}{2\sin(\Omega_1 t)}\exp(-\gamma t) - \cot(\Omega_1 t)\frac{\xi_{i1}+\xi_{i2}}{2}\right]\sin(\Omega_1\tau)\exp(\gamma\tau) +$$

$$+ \frac{\xi_{i1}+\xi_{i2}}{2}\cos(\Omega_1\tau)\exp(\gamma\tau) -$$

$$- \left[\frac{\xi_{f2}-\xi_{f1}}{2\sin(\Omega_2 t)}\exp(-\gamma t) - \cot(\Omega_2 t)\frac{\xi_{i2}-\xi_{i1}}{2}\right]\sin(\Omega_2\tau)\exp(\gamma\tau) -$$

$$- \frac{\xi_{i2}-\xi_{i1}}{2}\cos(\Omega_2\tau)\exp(\gamma\tau)$$

(A18)

$$\tilde{\xi}_2(\tau) = \left[\frac{\xi_{f1}+\xi_{f2}}{2\sin(\Omega_1 t)}\exp(-\gamma t) - \cot(\Omega_1 t)\frac{\xi_{i1}+\xi_{i2}}{2}\right]\sin(\Omega_1\tau)\exp(\gamma\tau) +$$

$$+ \frac{\xi_{i1}+\xi_{i2}}{2}\cos(\Omega_1\tau)\exp(\gamma\tau) +$$

$$+ \left[\frac{\xi_{f2}-\xi_{f1}}{2\sin(\Omega_2 t)}\exp(-\gamma t) - \cot(\Omega_2 t)\frac{\xi_{i2}-\xi_{i1}}{2}\right]\sin(\Omega_2\tau)\exp(\gamma\tau) +$$

$$+ \frac{\xi_{i2}-\xi_{i1}}{2}\cos(\Omega_2\tau)\exp(\gamma\tau)$$

(A19)

It is easy to verify from Eqs.(A16)-(A19) and Eqs.(A11),(A14) that in case of uncoupled oscillators ($\lambda = 0$) we have two independent trajectories identically coinciding with Eqs.(6.10),(6.11) from [15].

## Appendix B. Time-dependent functions for Eqs(6)-(8).

Here we wrote down the time dependent functions in equations for variances and covariance relating to the case of coupled identical oscillators.

For functions $e(t)$

$$e_3 = D_{12} + D'_{12} + \Pi_{15} - \frac{E_1(D'_4+\Pi_{16})}{2(C_2+\hbar a_2)}, \quad e_4 = \frac{(D_{11}+D'_{11}+\Pi_{14})(D'_4+\Pi_{16})}{2(C_2+\hbar a_2)},$$

$$e_5 = \frac{(D'_3+\Pi_8)(D'_4+\Pi_{16})}{2(C_2+\hbar a_2)}, \quad e_6 = \frac{(D_9+D'_9+\Pi_6)(D'_4+\Pi_{16})}{2(C_2+\hbar a_2)},$$

(B1)

For functions $Z(t)$

$$Z_1 = C_1/\hbar + a_1 - \frac{E_1^2}{4\hbar(C_2+\hbar a_2)} + \frac{e_3^2(C_2/\hbar+a_2)}{4\hbar a_2(C_2+\hbar a_2)+(D'_4+\Pi_{16})^2},$$

(B2)

$$Z_2 = D_3 + \Pi_5 - \frac{E_1(D_9 + D_9' + \Pi_6)}{2(C_2 + \hbar a_2)} - \frac{2e_3 e_6 (C_2 + \hbar a_2)}{4\hbar a_2 (C_2 + \hbar a_2) + (D_4' + \Pi_{16})^2},$$

(B3)

$$Z_3 = D_{10} + D_{10}' + \Pi_7 - \frac{E_1(D_3' + \Pi_8)}{2(C_2 + \hbar a_2)} - \frac{2e_3 e_5 (C_2 + \hbar a_2)}{4\hbar a_2 (C_2 + \hbar a_2) + (D_4' + \Pi_{16})^2}, \quad \text{(B4)}$$

$$Z_6 = D_4 + \Pi_{13} - \frac{E_1(D_{11} + D_{11}' + \Pi_{14})}{2(C_2 + \hbar a_2)} - \frac{2e_3 e_4 (C_2 + \hbar a_2)}{4\hbar a_2 (C_2 + \hbar a_2) + (D_4' + \Pi_{16})^2}, \quad \text{(B5)}$$

For functions $Y(t)$

$$Y_1 = a_1 + \frac{(D_{11} + D_{11}' + \Pi_{14})^2}{4\hbar(C_2 + \hbar a_2)} - \frac{e_4^2(C_2/\hbar + a_2)}{4\hbar a_2(C_2 + \hbar a_2) + (D_4' + \Pi_{16})^2} + \frac{Z_6^2}{4\hbar^2 Z_1}, \quad \text{(B6)}$$

$$Y_4 = \frac{i(D_3' + \Pi_8)(D_{11} + D_{11}' + \Pi_{14})}{2(C_2 + \hbar a_2)} - \frac{i2e_4 e_5(C_2 + \hbar a_2)}{4\hbar a_2(C_2 + \hbar a_2) + (D_4' + \Pi_{16})^2} + \frac{iZ_3 Z_6}{2\hbar Z_1}, \quad \text{(B7)}$$

$$Y_5 = \frac{i(D_9 + D_9' + \Pi_6)(D_{11} + D_{11}' + \Pi_{14})}{2(C_2 + \hbar a_2)} - \frac{i2e_4 e_6(C_2 + \hbar a_2)}{4\hbar a_2(C_2 + \hbar a_2) + (D_4' + \Pi_{16})^2} + \frac{iZ_2 Z_6}{2\hbar Z_1}, \quad \text{(B8)}$$

For functions $D(t)$

$$D_3(t) = (M/2)[-m_1 n_1(b_1 + b_{13}') + n_1(b_3 + b_{15}')/2 - m_2 n_2(b_{13} + b_1') + n_2(b_{15} + b_3')/2],$$
$$D_3'(t) = (M/2)[-m_1 \bar{n}_1(b_1 + b_{13}') + \bar{n}_1(b_3 + b_{15}')/2 - m_2 \bar{n}_2(b_{13} + b_1') + \bar{n}_2(b_{15} + b_3')/2],$$
$$D_4(t) = (M/2)[m_1^2(b_1 + b_{13}') - m_1(b_3 + b_{15}')/2 - m_1(b_2 + b_{14}')/2 + (b_4 + b_{16}')/4$$
$$+ m_2^2(b_1' + b_{13}) - m_2(b_3' + b_{15})/2 - m_2(b_2' + b_{14})/2 + (b_4' + b_{16})/4],$$
$$D_4'(t) = (M/2)[m_1^2(b_1 + b_{13}') - m_1(b_3 + b_{15}')/2 - m_1(b_2 + b_{14}')/2 + (b_4 + b_{16}')/4$$
$$+ m_2^2(b_1' + b_{13}) - m_2(b_3' + b_{15})/2 - m_2(b_2' + b_{14})/2 + (b_4' + b_{16})/4],$$
$$D_9(t) + D_9'(t) = (M/2)[-m_1 n_1(b_1 + b_{13}') + n_1(b_3 + b_{15}')/2 + m_2 n_2(b_{13} + b_1') - n_2(b_{15} + b_3')/2],$$
$$D_{10}(t) + D_{10}'(t) = (M/2)[-m_1 \bar{n}_1(b_1 + b_{13}') + \bar{n}_1(b_3 + b_{15}')/2 + m_2 \bar{n}_2(b_{13} + b_1') - \bar{n}_2(b_{15} + b_3')/2],$$
$$D_{11}(t) + D_{11}'(t) = (M/2)[m_1^2(b_1 + b_{13}') - m_1(b_3 + b_{15}')/2 - m_1(b_2 + b_{14}')/2 + (b_4 + b_{16}')/4 \quad \text{(B9)}$$
$$- m_2^2(b_1 + b_{13}') + m_2(b_{15} + b_3')/2 + m_2(b_{14} + b_2')/2 - (b_{16} + b_4')/4],$$
$$D_{12}(t) + D_{12}'(t) = (M/2)[m_1^2(b_1 + b_{13}') - m_1(b_3 + b_{15}')/2 - m_1(b_2 + b_{14}')/2 + (b_4 + b_{16}')/4$$
$$- m_2^2(b_1 + b_{13}') + m_2(b_{15} + b_3')/2 + m_2(b_{14} + b_2')/2 - (b_{16} + b_4')/4],$$

where

$$n_{1,2}(t) = \exp(\gamma t)/2\sin(\Omega_{1,2} t), \quad \bar{n}_{1,2}(t) = \exp(-\gamma t)/2\sin(\Omega_{1,2} t), \quad m_{1,2}(t) = \cot(\Omega_{1,2} t)/2, \quad \text{(B10)}$$

It should be noted that in case identical oscillators we have from (B9) that $D_3 = D_3'$, $D_4 = D_4'$, $D_9 + D_9' = D_{10} + D_{10}'$, $D_{11} + D_{11}' = D_{12} + D_{12}'$.

$$b_1 = \Omega_1^2 s_1 - (\omega_0^2 - \gamma^2)s_2 - 2\Omega_1\gamma s_5, b_2 = -\Omega_1^2 s_5 - (\omega_0^2 - \gamma^2)s_5 - \Omega_1\gamma(s_1 - s_2),$$
$$b_3 = -\Omega_1^2 s_5 - (\omega_0^2 - \gamma^2)s_5 - \Omega_1\gamma(s_1 - s_2), b_4 = \Omega_1^2 s_2 - (\omega_0^2 - \gamma^2)s_1 + 2\Omega_1\gamma s_5,$$
$$b_5 = -\Omega_1\Omega_2 s_7 + (\omega_0^2 - \gamma^2)s_{10} + \Omega_1\gamma s_8 + \Omega_2\gamma s_9, b_6 = \Omega_1\Omega_2 s_8 + (\omega_0^2 - \gamma^2)s_9 + \Omega_1\gamma s_7 - \Omega_2\gamma s_{10},$$
$$b_7 = \Omega_1\Omega_2 s_9 + (\omega_0^2 - \gamma^2)s_8 - \Omega_1\gamma s_{10} + \Omega_2\gamma s_7, b_8 = -\Omega_1\Omega_2 s_{10} + (\omega_0^2 - \gamma^2)s_7 - \Omega_1\gamma s_9 - \Omega_2\gamma s_8,$$
$$b_9 = -\Omega_1\Omega_2 s_{11} + (\omega_0^2 - \gamma^2)s_{14} + \Omega_2\gamma s_{12} + \Omega_1\gamma s_{13}, b_{10} = \Omega_1\Omega_2 s_{12} + (\omega_0^2 - \gamma^2)s_{13} + \Omega_2\gamma s_{11} - \Omega_1\gamma s_{14}, \quad (B11)$$
$$b_{11} = \Omega_1\Omega_2 s_{13} + (\omega_0^2 - \gamma^2)s_{12} - \Omega_2\gamma s_{14} + \Omega_1\gamma s_{11}, b_{12} = -\Omega_1\Omega_2 s_{14} + (\omega_0^2 - \gamma^2)s_{11} - \Omega_2\gamma s_{13} - \Omega_1\gamma s_{12},$$
$$b_{13} = \Omega_2^2 s_3 - (\omega_0^2 - \gamma^2)s_4 - 2\Omega_2\gamma s_6, b_{14} = -\Omega_2^2 s_6 - (\omega_0^2 - \gamma^2)s_6 + \Omega_2\gamma(s_4 - s_3),$$
$$b_{15} = -\Omega_2^2 s_6 - (\omega_0^2 - \gamma^2)s_6 + \Omega_2\gamma(s_4 - s_3), b_{16} = \Omega_2^2 s_4 - (\omega_0^2 - \gamma^2)s_3 + 2\Omega_2\gamma s_6,$$

$$b_1' = \Omega_2^2 s_3 - (\omega_0^2 - \gamma^2)s_4 - 2\Omega_2\gamma s_6, b_2' = -\Omega_2^2 s_6 - (\omega_0^2 - \gamma^2)s_6 - \Omega_2\gamma(s_3 - s_4),$$
$$b_3' = -\Omega_2^2 s_6 - (\omega_0^2 - \gamma^2)s_6 - \Omega_2\gamma(s_3 - s_4), b_4' = \Omega_2^2 s_4 - (\omega_0^2 - \gamma^2)s_3 + 2\Omega_2\gamma s_6,$$
$$b_5' = \Omega_1\Omega_2 s_7 - (\omega_0^2 - \gamma^2)s_{10} - \Omega_1\gamma s_8 - \Omega_2\gamma s_9, b_6' = -\Omega_1\Omega_2 s_8 - (\omega_0^2 - \gamma^2)s_9 - \Omega_1\gamma s_7 + \Omega_2\gamma s_{10},$$
$$b_7' = \Omega_1\Omega_2 s_{10} - (\omega_0^2 - \gamma^2)s_7 + \Omega_1\gamma s_9 + \Omega_2\gamma s_8, b_8' = -\Omega_1\Omega_2 s_9 - (\omega_0^2 - \gamma^2)s_8 + \Omega_1\gamma s_{10} - \Omega_2\gamma s_7,$$
$$b_9' = \Omega_1\Omega_2 s_{11} - (\omega_0^2 - \gamma^2)s_{14} - \Omega_2\gamma s_{12} - \Omega_1\gamma s_{13}, b_{10}' = -\Omega_1\Omega_2 s_{12} - (\omega_0^2 - \gamma^2)s_{13} - \Omega_2\gamma s_{11} + \Omega_1\gamma s_{14}, \quad (B12)$$
$$b_{11}' = -\Omega_1\Omega_2 s_{13} - (\omega_0^2 - \gamma^2)s_{12} + \Omega_2\gamma s_{14} - \Omega_1\gamma s_{11}, b_{12}' = \Omega_1\Omega_2 s_{14} - (\omega_0^2 - \gamma^2)s_{11} + \Omega_2\gamma s_{13} + \Omega_1\gamma s_{12},$$
$$b_{13}' = \Omega_1^2 s_1 - (\omega_0^2 - \gamma^2)s_2 - 2\Omega_1\gamma s_5, b_{14}' = -\Omega_1^2 s_5 - (\omega_0^2 - \gamma^2)s_5 + \Omega_1\gamma(s_2 - s_1),$$
$$b_{15}' = -\Omega_1^2 s_5 - (\omega_0^2 - \gamma^2)s_5 + \Omega_1\gamma(s_2 - s_1), b_{16}' = \Omega_1^2 s_2 - (\omega_0^2 - \gamma^2)s_1 + 2\Omega_1\gamma s_5,$$

where

$$s_1(t) = t/2 + \sin(2\Omega_1 t)/4\Omega_1, \quad s_2(t) = t/2 - \sin(2\Omega_1 t)/4\Omega_1, \quad s_3(t) = t/2 + \sin(2\Omega_2 t)/4\Omega_2,$$
$$s_4(t) = t/2 - \sin(2\Omega_2 t)/4\Omega_2, \quad s_5(t) = \sin^2(\Omega_1 t)/2\Omega_1, \quad s_6(t) = \sin^2(\Omega_2 t)/2\Omega_2, \quad (B13)$$

$$s_7(t) = s_{11}(t) = \frac{\Omega_1 \cos(t\Omega_2)\sin(t\Omega_1) - \Omega_2 \cos(t\Omega_1)\sin(t\Omega_2)}{\Omega_1^2 - \Omega_2^2}, \quad (B14)$$

$$s_8(t) = s_{13}(t) = \frac{-\Omega_2 + \Omega_2 \cos(t\Omega_2)\cos(t\Omega_1) + \Omega_1 \sin(t\Omega_1)\sin(t\Omega_2)}{\Omega_1^2 - \Omega_2^2}, \quad (B15)$$

$$s_9(t) = s_{12}(t) = \frac{\Omega_1 - \Omega_1 \cos(t\Omega_2)\cos(t\Omega_1) - \Omega_2 \sin(t\Omega_1)\sin(t\Omega_2)}{\Omega_1^2 - \Omega_2^2}, \quad (B16)$$

$$s_{10}(t) = s_{14}(t) = \frac{\Omega_2 \cos(t\Omega_2)\sin(t\Omega_1) - \Omega_1 \cos(t\Omega_1)\sin(t\Omega_2)}{\Omega_1^2 - \Omega_2^2}. \quad (B17)$$

We note that due to (B14)-(B17) it is follows from (B11)-(B12) that $b_5 = -b_5'$, $b_6 = -b_6'$, $b_7 = -b_8'$, $b_8 = -b_7'$, $b_9 = -b_9'$, $b_{10} = -b_{10}'$, $b_{11} = -b_{11}'$, $b_{12} = -b_{12}'$.

For functions $\Pi(t)$

$$\Pi_5(t) = \Pi_8(t) = \lambda(-n_1 m_1 s_2 + n_1 s_5/2 + n_2 m_2 s_4 - n_2 s_6/2),$$
$$\Pi_6(t) = \Pi_7(t) = \lambda(-n_1 m_1 s_2 + n_1 s_5/2 - n_2 m_2 s_4 + n_2 s_6/2),$$
$$\Pi_{13}(t) = \Pi_{16}(t) = \lambda(m_1^2 s_2 - m_1 s_5 + s_1/4 - m_2^2 s_4 + m_2 s_6 - s_3/4), \quad (B18)$$
$$\Pi_{14}(t) = \Pi_{15}(t) = \lambda(m_1^2 s_2 - m_1 s_5 + s_1/4 + m_2^2 s_4 - m_2 s_6 + s_3/4),$$

The functions $C_{1,2}$, $E_1$ designated as $R_k$ ($k=1,2$) can be represented as follows

$$R_k(t) = \frac{2M\gamma}{\pi}\left\{\int_0^\infty d\omega\,\omega\,Coth\left(\frac{\hbar\omega}{2k_B T_1}\right)\int_0^t\int_0^\tau ds\,d\tau\, R_k^{(1)}(\tau,s)\cos[\omega(\tau-s)]\exp[\gamma(\tau+s)]\right.$$
$$\left.+\int_0^\infty d\omega\,\omega\,Coth\left(\frac{\hbar\omega}{2k_B T_2}\right)\int_0^t\int_0^\tau ds\,d\tau\, R_k^{(2)}(\tau,s)\cos[\omega(\tau-s)]\exp[\gamma(\tau+s)]\right\}, \quad (B19)$$

where

$$C_1^{(1)}(\tau,s) = C_2^{(2)}(\tau,s) = m_1^2 f_1 + m_2^2 f_2 + m_1 m_2(f_3+f_4)$$
$$-m_1(f_9+f_{10}+f_{11}+f_{16})/2 - m_2(f_{12}+f_{15}+f_{13}+f_{14})/2 + (f_5+f_6+f_7+f_8)/4], \quad (B20)$$

$$C_1^{(2)}(\tau,s) = C_2^{(1)}(\tau,s) = m_1^2 f_1 + m_2^2 f_2 - m_1 m_2(f_3+f_4)$$
$$+m_1(f_{11}+f_{16}-f_9-f_{10})/2 + m_2(f_{12}+f_{15}-f_{13}-f_{14})/2 + (f_5+f_6-f_7-f_8)/4]$$

$$E_1^{(1)}(\tau,s) = E_1^{(2)}(\tau,s) = 2m_1^2 f_1 - 2m_2^2 f_2 - m_1(f_9+f_{10}) + m_2(f_{13}+f_{14}) + (f_5-f_6)/2] \quad (B21)$$

where all $f_i$, ($i=1,...,16$) are the functions of $\tau$ and $s$:

$$f_1(\tau,s) = sin(\Omega_1\tau)sin(\Omega_1 s),\ f_2(\tau,s) = sin(\Omega_2\tau)sin(\Omega_2 s),$$
$$f_3(\tau,s) = sin(\Omega_1\tau)sin(\Omega_2 s),\ f_4(\tau,s) = sin(\Omega_2\tau)sin(\Omega_1 s),$$
$$f_5(\tau,s) = \cos(\Omega_1\tau)\cos(\Omega_1 s),\ f_6(\tau,s) = \cos(\Omega_2\tau)\cos(\Omega_2 s),$$
$$f_7(\tau,s) = \cos(\Omega_1\tau)\cos(\Omega_2 s),\ f_8(\tau,s) = \cos(\Omega_2\tau)\cos(\Omega_1 s), \quad (B22)$$
$$f_9(\tau,s) = sin(\Omega_1\tau)\cos(\Omega_1 s),\ f_{10}(\tau,s) = \cos(\Omega_1\tau)sin(\Omega_1 s),$$
$$f_{11}(\tau,s) = sin(\Omega_1\tau)\cos(\Omega_2 s),\ f_{12}(\tau,s) = \cos(\Omega_1\tau)sin(\Omega_2 s),$$
$$f_{13}(\tau,s) = sin(\Omega_2\tau)\cos(\Omega_2 s),\ f_{14}(\tau,s) = \cos(\Omega_2\tau)sin(\Omega_1 s),$$
$$f_{15}(\tau,s) = sin(\Omega_2\tau)\cos(\Omega_1 s),\ f_{16}(\tau,s) = \cos(\Omega_2\tau)sin(\Omega_1 s),$$

It should be emphasized that the integration with respect to $\tau$ and $s$ in Eq.(B19) can be performed analytically, but the final results are very cumbersome and we wrote down here only integral forms. Nevertheless, we have done it in our numerical calculations in order to shorten considerably a computer consuming time.

**Figure captions:**

**Figure 1.** Temporal dynamics of normalized variances $\sigma_{1,2}^2(t)/\sigma_{1,2}^2(FDT)$ for the first (curve 1) and second (curve 2) coupled oscillators interacting with separate thermostats at different temperatures $T_1 = 300K$ and $T_2 = 700K$ in accordance with Eqs.(10), (11) and (16). Initial variances $\sigma_{01,02}^2 = \hbar/2M\omega_0$ correspond to the natural case of the cold system at $T_1 = T_2 = 0K$. The normalized coupling constant $\tilde{\lambda} = \lambda/M\omega_0^2$ is shown in the insets of each picture. Normalization was done to the variances of oscillators in equilibrium $\sigma_1^2(FDT)$ at $T_1 = 300K$ and $\sigma_2^2(FDT)$ at $T_2 = 700K$ correspondingly.

**Figure 2.** Temporal dynamics of normalized variances $\sigma_{1,2}^2(t)/\sigma_{1,2}^2(FDT)$ for the first (curve 1) and second (curve 2) coupled oscillators as well as in figure 1, but with initial variances $\sigma_{01}^2 = 10(\hbar/2M\omega_0)$ and $\sigma_{02}^2 = \hbar/2M\omega_0$ relating to the arbitrary initial state of the first oscillator. The normalized coupling constant $\tilde{\lambda} = \lambda/M\omega_0^2$ is shown in the insets of each picture.

**Figure 3.** Normalized variances $\tilde{\sigma}_{1,2}^2(t = \infty)$ (curves 1, 2) and covariance $\tilde{\beta}_{12}^{-1}(t = \infty)$ (curve 3) of coupled oscillators interacting with different thermostats versus the normalized coupling constant $\tilde{\lambda}$. The separate thermostats are kept with identical temperatures $T_1 = T_2 = 300K$ -a), and with different temperatures $T_1 = 300K$ and $T_2 = 700K$ -b). In figure 3c) the covariances $\tilde{\beta}_{12}^{-1}(t = \infty)$ are presented in case of equal temperatures $T_1 = T_2 = 300K$ (low curve) and at different temperatures $T_1 = 300K$ and $T_2 = 700K$ (upper curve) for small values of coupling constants. Calculations were done with use of Eqs.(10), (11) and (16).

**Figure 4.** Sketches illustrating the equivalence of Hamiltonians in Eq.(17) (upper sketch) and (19) (low sketch) at $\lambda = M\omega_0^2$.

**Figure 5.** Temperature dependencies of normalized variances $\tilde{\sigma}_{1,2}^2(t = \infty)$ in quasistationary states versus $T_2$ at fixed $T_1 = 300K$. The dashed curves 1, 2 at $\tilde{\lambda} = 0.01$ and continuous curves 1, 2 at $\tilde{\lambda} = 0.1$.